\documentclass[prb,twocolumn,amsmath,amssymb,preprintnumbers,superscriptaddress]{revtex4}
\usepackage{dcolumn}
\usepackage{bm}
\usepackage{graphicx}
\usepackage{color}

\begin{document}

\title{Current-voltage characteristics and resistive switching in epitaxial La$_{0.67}$Sr$_{0.33}$MnO$_3$/SrMnO$_3$/La$_{0.67}$Sr$_{0.33}$MnO$_3$ multilayer}

\author{A. G. A. Rahman}\affiliation{School of Physical Sciences, Jawaharlal Nehru University, New Delhi - 110067, India.}
\author{R. K. Patel}\affiliation{School of Physical Sciences, Jawaharlal Nehru University, New Delhi - 110067, India.}
\author{Chandrani Nath}\affiliation{Special Centre for Nanoscience, Jawaharlal Nehru University, New Delhi - 110067, India.}
\author{A. K. Pramanik}\email{akpramanik@mail.jnu.ac.in}\affiliation{School of Physical Sciences, Jawaharlal Nehru University, New Delhi - 110067, India.}

\maketitle

\onecolumngrid

\vspace{-6mm}
\section*{Abstract}
In present study, we investigate the structural properties and current-voltage ($I$-$V$) characteristics of an epitaxial multilayer composed of La$_{0.67}$Sr$_{0.33}$MnO$_3$ (25 nm)/SrMnO$_3$ (70 nm)/La$_{0.67}$Sr$_{0.33}$MnO$_3$ (25 nm), grown on different substrates i.e., SrTiO$_3$(100), LaAlO$_3$(100), Si (100). The used substrates not only have different chemical compositions in line with the film materials, also they have different lattice parameters which would tune the material chemistry and lattice strain at the interface significantly. Our $I$-$V$ measurements reveal distinct electrical behaviors depending on the substrate. The multilayers grown on oxide SrTiO$_3$(100) and LaAlO$_3$(100) substrates exhibit regular $I$-$V$ with slight nonlinearity at low applied voltages. In contrast, $I$-$V$ in multilayer with Si(100) substrate exhibits large asymmetry and notable resistive switching (RS) behavior at room temperature, where a resistance ratio around 10 between the high resistance state (HRS) and low resistance state (LRS) is observed. While the observed $I$-$V$ behavior is sensitive to substrate-induced interfacial disorder and formation of trap bands within bandgap of SrMnO$_3$, the present results have potential applications in future memory and memristive devices.            

\vspace{2mm}
\noindent\textbf{Keywords:} Oxide heterostructure, Current-voltage characteristics, Resistive switching, Interfacial strain, Space charge limited conduction

\vspace{3mm}
\noindent\textbf{pacs:} 79.60.Jv, 73.50.-h, 73.40.-c, 78.30.-j

\vspace{5mm} 

\twocolumngrid

\section{Introduction}
Phenomenon of resistive switching (RS) has been the key area of research for last several decades, when this RS in association with voltage-controlled negative resistance behavior laid the foundation stone for today's non-volatile memory technologies.\cite{hiatt} The RS characterizes a transition between two stable states of resistance i.e., high-resistance states (HRS) and low-resistance states (LRS) in response to electrical inputs, and has been prominently used in devices of resistive random-access memory (RRAM). A simple structure of metal-insulator-metal (MIM) constitutes the working unit of RRAM, therefore owing to its high density, fast switching, high retention, and low power consumption the RRAM is considered to be the most promising candidate for next-generation memory devices.\cite{waser,shang,pan} 

Till so far, the RS behavior has been explained with different microscopic models related to redox-process driven ion migration,\cite{baikalov,rwaser} diffusion of oxygen vacancies,\cite{nian} formation/rapture of conducting filaments in intermediate insulating layer,\cite{liu,chudnovskii,rossel,kwon,lee} formation of interfacial metal-oxide layer and electron dynamics,\cite{tsui,odagawa} etc. Nonetheless, RS has widely been classified into two types (i.e., unipolar and bipolar) based on the effect of voltage polarity on switching behavior. Unipolar switching between LRS and HRS occurs with the same voltage polarity but having different magnitudes, while the bipolar switching between resistance states requires opposite voltage polarities. The RS has further been categorized based on its conduction mechanisms i.e., whether the formation and rupture of conducting filaments (CFs) occurs within the insulating matrix or at the interface between the electrode and the insulator.\cite{sawa} A wide class of materials including oxides, chalcogenides, carbon-based materials, nitrides, and organic compounds are shown to exhibit RS behavior,\cite{gao,szot,lai,guo,zhou,chen,kim} however, oxides are particularly promising due to their rich functionality added with wide range of composition and electronic properties. The transport of oxygen vacancies as well as metal ions largely characterizes the RS behavior in oxides, and these materials are considered leading candidates for the development of next-generation memory devices.\cite{khare,sawa,kubicek,hu} While a significant progress has already been achieved, further studies using different kinds of electrodes and substrate strain are required for a detailed understanding of RS behavior.

\begin{figure*}
	\centering
		\includegraphics[width=16cm]{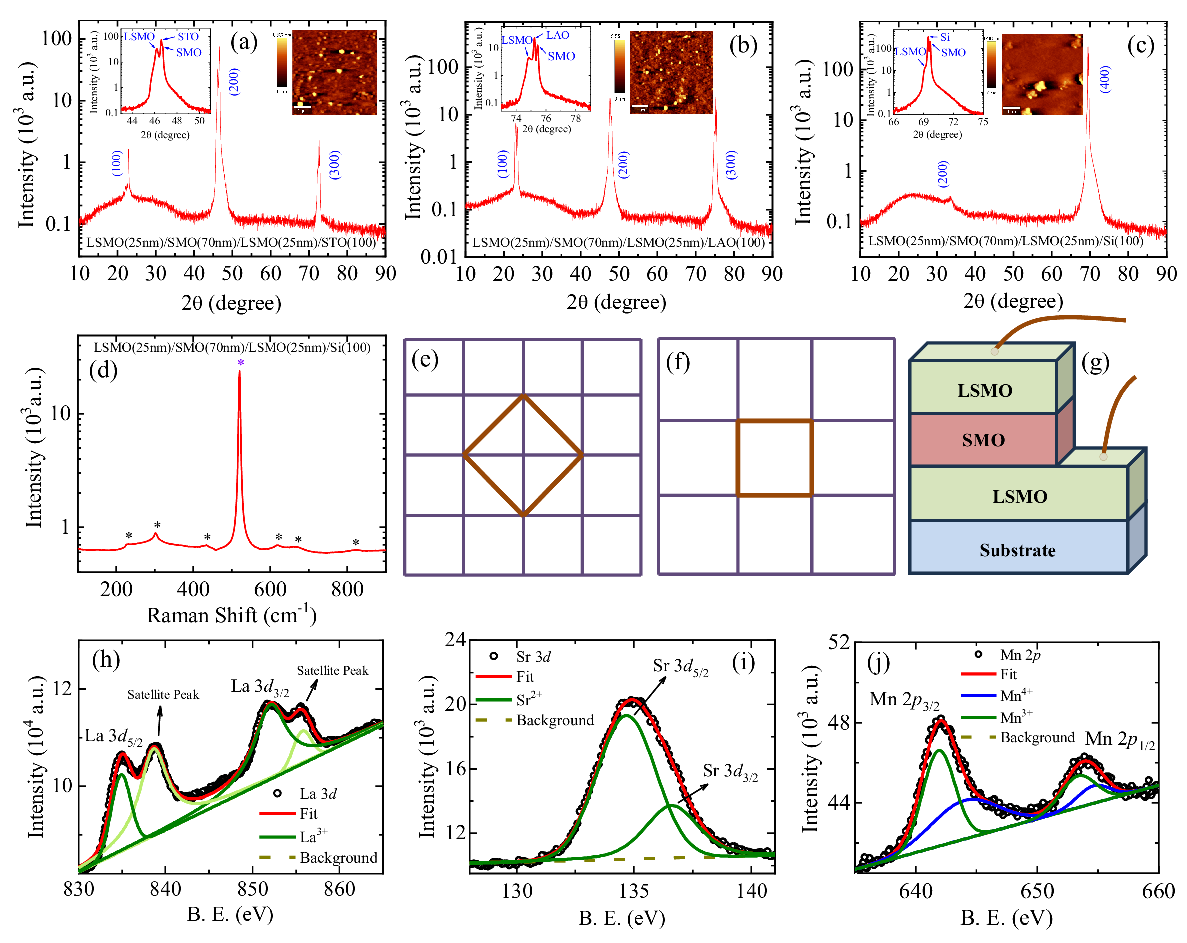}
	\caption{(a),(b) and (c) show the room temperature XRD data of LSMO/SMO/LSMO trilayer deposited on STO(100), LAO(100) and Si(100) substrates in semi-log scale, respectively. The left inset of (a), (b), and (c) shows a magnified view of the (200), (300), and (400) peaks of the STO, LAO, and Si based films, respectively. The right inset displays the AFM images of respective film. (d) shows the Raman spectra of Si-based film at room temperature. (e) and (f) show the schematic of film-growth orientation with substrate while (g) shows the schematic of layer arrangement in present multilayer with electrical connections for $I$-$V$ measurements. (h), (i) and (j) show the x-ray photoemission spectroscopy (XPS) spectra related to La-3$d$, Sr-3$d$ and Mn-2$p$, respectively where the open black circles represent the experimental data and the red solid lines are due to fitted envelope. Other solid lines are due to different cationic charge states, as specified in the plot.}
	\label{fig:Fig1}
\end{figure*}

In this study, we have investigated the current-voltage ($I$-$V$) properties in a multilayer (ML) oxide film consisting of La$_{0.67}$Sr$_{0.33}$MnO$_3$ (25 nm)/SrMnO$_3$ (70 nm)/La$_{0.67}$Sr$_{0.33}$MnO$_3$ (25 nm). The La$_{1-x}$Sr$_{x}$MnO$_3$ is an interesting series of materials showing various electronic and magnetic phases with $x$.\cite{urushibara, hemberger} The two end compositions (i.e., $x$ = 0.0 and 1.0) are antiferromagnetic insulator (AFM-I) but the intermediate composition with $x$ = 0.33 shows metallic (M) and ferromagnetic (FM) behavior with highest ordering temperature $T_c$ $\sim$ 370 K. On stoichiometric consideration, La$_{0.67}$Sr$_{0.33}$MnO$_3$ (LSMO) has mixed Mn$^{3+}$ (3$d^4$) and Mn$^{4+}$ (3$d^3$) valance while SrMnO$_3$ (SMO) has only Mn$^{4+}$, so this trilayer consists of materials with different electron densities or conductivities giving it a M-I-M configuration. Further, in addition to working as electrodes, LSMO acts as reservoir for oxygen vacancies and/or electrons due to its different conductivity with SMO. A recent study has shown a better control on RS properties where a material combination with different conductivities (Ta$_2$O$_{5-x}$/TaO$_{2-x}$) has been used between Pt electrodes.\cite{mjlee} Structurally, LSMO has rhombohedral-\textit{R$\bar{3}$c} symmetry with lattice paramater $a$ $\sim$ 5.461 {\AA}).\cite{urushibara} The SMO, on the other hand, has three polymorphs: 4$H$-hexagonal, 6$H$-hexagonal and cubic ($C$) perovskite, however, the most commonly occurring SMO phase is 4$H$-hexagonal (space group \textit{P6$_3$/mmc}, lattice parameters $a$ = $b$ = 5.454 {\AA} and $c$ = 9.092 {\AA}) with AFM transition temperature $T_N$ $\sim$ 280 K.\cite{hemberger,ulrichs} Due to proximity in lattice parameters, an epitaxial LSMO/SMO/LSMO trilayer is expected in $c$-axis (001) orientation. Further, to understand the effect of underlying lattice strain, the films have been grown on different substrates such as, SrTiO$_3$ (100), LaAlO$_3$ (100) and Si (100), using pulsed laser deposition (PLD) technique. Our $I-V$ results show slight nonlinearity on oxide substrates, but a clear RS behavior has been observed in film with Si (100) substrate. These behavioral differences underline the importance of lattice strain in controlling the complicated oxide thin film for electrical performances.

\begin{figure*}
	\centering
		\includegraphics[width=16cm]{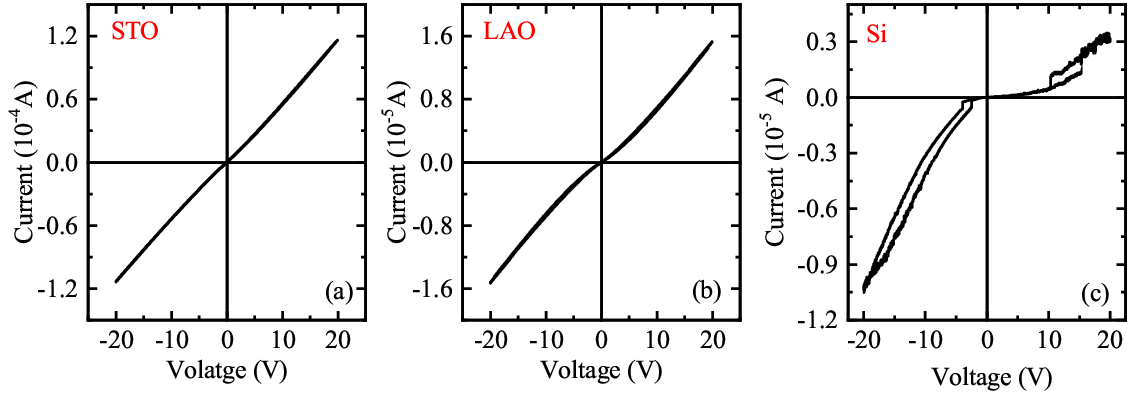}
	\caption{The $I$-$V$ characteristics collected at room temperature with voltage sweeping range $\pm$20 V are shown for LSMO/SMO/LSMO multilayer with (a) STO, (b) LAO and (c) Si substrate.}
	\label{fig:Fig2}
\end{figure*}

\section{Experimental Details}   
Trilayer epitaxial films composed of La$_{0.67}$Sr$_{0.33}$MnO$_3$ (25 nm)/SrMnO$_3$ (70 nm)/La$_{0.67}$Sr$_{0.33}$MnO$_3$ (25 nm) are deposited on single-crystal substrates of SrTiO$_3$ (100), LaAlO$_3$ (100) and Si (100) using the pulsed laser deposition (PLD) method, employing a KrF laser with wavelength of 248 nm. The LSMO and SMO targets have been prepared using solid-state reaction methods. The LSMO and SMO targets have been synthesized by mixing stoichiometric ratio of powder ingredients La$_2$CO$_3$, SrCO$_3$ and MnO$_2$. The rare-earth oxide has been given preheat treatment at 1000$^{\circ}$C for 10 hours to remove absorbed moisture. The mixed powders are ground well and calcined at temperature 900$^{\circ}$C for 24 hours. Then, the powders are pressed into pellets and given heat treatment at high temperatures with intermediate grindings, where the final treatment for LSMO and SMO has been given at 1200 and 1100$^{\circ}$C for 24 hours. The crystalline phase for both targets is checked with x-ray diffraction (XRD) measurements. For film deposition, the substrate temperature, the oxygen pressure, and the pulse frequency are used as 700$^{\circ}$C, 0.05 mbar and 5 Hz, respectively. Throughout the deposition, the laser energy density at the target surface is fixed at $\sim$ 1.3 J/cm$^2$. An approximate film thickness is estimated via laser shot calibration.\cite{kharkwal} For $I$-$V$ measurements, an initial $\sim$ 25 nm thick LSMO film is deposited on the substrates (STO, LAO, Si). After cooling to room temperature, a small portion of the deposited LSMO film has been masked, and then subsequent layers of $\sim$70 nm SMO and $\sim$25 nm LSMO have been grown under the same conditions (see Fig. 1(g)). The top and bottom LSMO layers, known for their metallic properties, serve as electrodes.

The structural properties of the films are evaluated using X-ray diffraction (XRD). The surface quality of the films is examined using atomic force microscopy (AFM) at room temperature. The Raman measurements have been conducted using using a WITec alpha300 microscope (Oxford) with a 532 nm excitation wavelength. The $I$-$V$ characteristics have been measured by scanning the applied voltage and recording the response current using a Keithley-2400 source meter with two-terminal geometry where the electrical contacts are made on top and bottom LSMO electrodes. The sign of the voltages in plots corresponds to the voltages applied on the top LSMO layer (Fig. 1(g)). The x-ray photoemission spectroscopy (XPS) measurements have been done with a lab-based spectrometer (SPECS, Germany) using Al-$K_{\alpha}$ (1486.61 eV) x-ray source. The XPS data are analyzed with a XPSPeak4.1 software.

\begin{figure*}
	\centering
		\includegraphics[width=16cm]{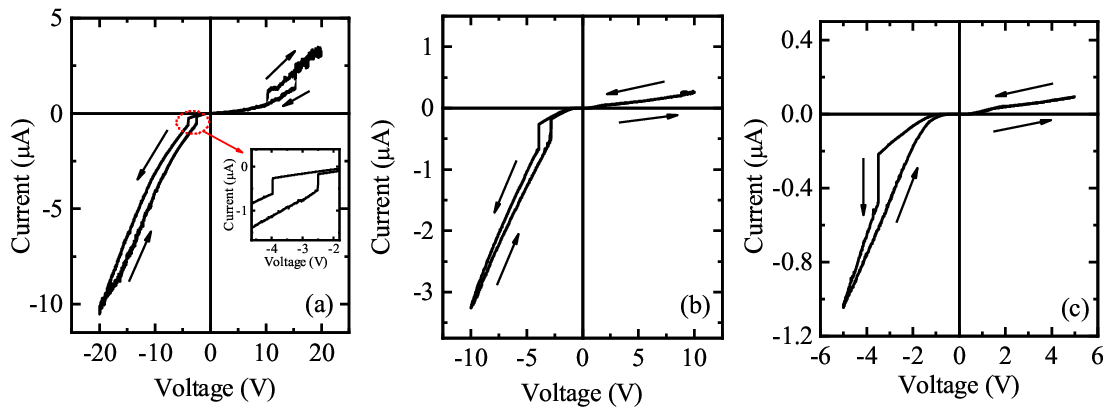}
	\caption{(a), (b) and (c) show the $I$-$V$ characteristics of LSMO/SMO/LSMO film deposited on Si(100) substrate at voltage sweeping range $\pm$20 V, $\pm$10 V and $\pm$5 V, respectively. Arrows show the direction of voltage cycle. Inset of (a) presents the magnified view of $I(V)$ data in negative low voltage regime showing a clear RS behavior.}
	\label{fig:Fig3}
\end{figure*}

\section{Results and Discussion}
\subsection{Structural and chemical analysis}
Figures 1(a), 1(b), and 1(c) present the $\theta$--$2\theta$ x-ray diffraction (XRD) plots for LSMO/SMO/LSMO films grown on STO, LAO, and Si substrates, respectively each with a thickness of approximately 120 nm. The XRD data clearly demonstrate that the films possess a single-crystalline nature and maintain an epitaxial relationship with all three substrates. The AFM images as shown in right inset of Fig. 1(a), (b) and (c) shows smooth surface with small roughness around 1.40 nm, 0.72 nm and 1.90 nm for STO, LAO and Si-based films respectively.

The left insets of Figs. 1(a), (b) and (c) show a magnified view of the (200),(300) and (400) diffraction peaks of the deposited materials along with respective substrates. In all cases, the Bragg peaks due to LSMO and SMO appear on the left and right side of the substrate peak, respectively. From the XRD plot in Fig. 1(a), we have calculated the pseudo-cubic lattice parameters of the film with $a_{LSMO}$ = 3.92 {\AA}, $a_{SMO}$ = 3.88 {\AA} and for the STO substrate $a_{STO}$= 3.89 {\AA}. Similarly from Fig. 1(b), the calculated pseudo-cubic lattice parameters of the LAO film are $a_{LSMO}$ = 3.80 {\AA}, $a_{SMO}$ = 3.77 {\AA} and for the LAO substrate $a_{LAO}$ = 3.78 {\AA}. The STO and LAO substrates have cubic structure with lattice parameter $a_{STO}$= 3.90 {\AA} and $a_{LAO}$= 3.82 {\AA}, which suggest an epitaxial growth of LSMO and SMO on these substrates, given that $a_{LSMO/SMO}$ $\sim$ $\sqrt{2}a_{STO/LAO}$. In both cases, the growth of LSMO/SMO films on STO/LAO (100) would be 45$^{\circ}$ rotated, as shown in Fig. 1(e). From XRD (Figs. 1(a) and (b)), we have calculated the lattice strain for LSMO/SMO on STO substrate are $\sim$ -1.2\%/-0.77\% while on LAO the values are $\sim$ +1.8\%/+3.77\%, respectively. Similarly, from the XRD pattern of Si-based film (Fig. 1(c)), the calculated pseudo-cubic lattice parameters are $a_{LSMO}$ = 5.43 {\AA}, $a_{SMO}$= 5.41 {\AA} and for the Si substrate $a_{Si}$= 5.40 {\AA}. The calculated lattice strain for LSMO/SMO on Si film are $\sim$ +0.73\%/+0.73\%, respectively. In this case, the growth of films on Si substrate is 0$^{\circ}$ rotated, as shown in Fig. 1(f)).

To further verify the structural phase, room temperature Raman spectroscopy measurements have been done for Si based film, as presented in Fig. 1(d). There are total seven broad or weak Raman peaks are seen at 231, 302, 434, 520, 617, 666 and 823 cm$^{-1}$, where the peak at 520 cm$^{-1}$ is due to Si substrate peak. The LSMO and SMO have commonality of MnO$_6$ octahedra, but they differ in lattice structure and their Mn$^{3+}$/Mn$^{4+}$ charge states. In fact, the electronic and magnetic properties in La$_{1-x}$Sr$_x$MnO$_3$ show a large dependence on La/Sr content. \cite{urushibara} The group theoretical analysis suggests five active Raman modes ($A_{1g}$ + 4$E_g$) for LSMO (rhombohedral-\textit{R$\bar{3}$c}) and eight active Raman modes (2$A_{1g}$ + 2$E_{1g}$ + 4$E_{2g}$) for hexagonal SMO (\textit{P6$_3$/mmc}), however, no Raman active phonon has been predicted for cubic SMO (\textit{Pm3m}) phase.\cite{granado,sacchetti} While bulk single-crystal LSMO has shown only two Raman active peaks around 199 and 426 cm$^{-1}$, \cite{granado} the surface and interface lattice strain in films has dominant role on the Raman peaks, even modifying the peak position with layer thickness as well as with other materials in heterostructures.\cite{behera,rlijin,anne,ren,aga} Nonetheless, the Raman modes in these materials are broadly classified into two distinct frequency regimes: the low-frequency regime (below $\sim$ 300 cm$^{-1}$) is related to the bending/rotation of MnO$_6$ octahedra while the high-frequency regime (above $\sim$ 500 cm$^{-1}$) is linked to stretching of Mn-O bonds. Nonetheless, the observed modes are due to combination of LSMO and SMO phases. Our Raman results further show that SMO adopts hexagonal phase in present trilayer.

To understand the chemical composition in present ML, we have performed x-ray photoemission spectroscopy (XPS) measurements at room temperature. Here, we mention that unlike Raman spectroscopy, XPS is extremely surface sensitive and can probe only limited atomic layers beneath the surface. Given that our ML composition is LSMO(25 nm)/SMO(70 nm)/LSMO(25 nm), therefore it is unlikely that the sandwiched insulating SMO layer can be studied with the XPS measurements. Nonetheless, both LSMO ($x$ = 0.33) and SMO ($x$ = 1) belong to same La$_{1-x}$Sr$_x$MnO$_3$ composition, and their bulk targets as well as films are prepared following similar techniques. Therefore, XPS will mostly investigate the top LSMO, while we expect there would be some ionic contributions from SMO as well. Figs. 1(h), 1(i) and 1(j) present the measured XPS spectrum (open black circles) related to La-3$d$, Sr-3$d$ and Mn-2$p$ level, respectively for present ML film, while the solid red lines in figures are due to overall fitting of the data. Fig. 1(h) shows spin-orbit split two peaks at the binding energies (BE) 834.88 and 852.05 eV (solid green line) due to La-3$d_{5/2}$ and La-3$d_{3/2}$, respectively confirming a La$^{3+}$ electronic state.\cite{yangconf,xpsbook} The figure shows additional two La-3$d$ satellite peaks at BE 838.62 and 855.71 eV.\cite{burroughs,viswanathan,vasquez} A similar spin-orbit split two peaks (solid green lines) due to Sr-3$d_{5/2}$ and Sr-3$d_{3/2}$ are observed at BE 134.68 and 136.61 eV, respectively in Fig. 1(i).\cite{yangconf,bertaccor} This indicates Sr$^{2+}$ electronic state in present ML. The analysis of Mn-2$p$ XPS spectra in Fig. 1(j) suggests mixed Mn$^{3+}$ and Mn$^{4+}$ valency, as represented by green and blue solid lines, respectively. For each charge state, spin-orbit split two peaks due to Mn-2$p_{3/2}$ and Mn-2$p_{1/2}$ are observed. For Mn-2$p_{3/2}$, the peak positions of Mn$^{3+}$ and Mn$^{4+}$ are observed at BE 641.89 and 644.01 eV, while for for Mn-2$p_{1/2}$ the same are observed at 653.37 and 654.62 eV, respectively.\cite{yangconf,chaluvadi,xieh} Our analysis implies the Mn$^{3+}$/Mn$^{4+}$ ratio is about 57/43. While for stoichiometric LSMO (top layer), the Mn$^{3+}$/Mn$^{4+}$ ratio is expected to be 67/33, but the observed higher amount of Mn$^{4+}$ suggests a contribution from intermediate SMO layer which has almost Mn$^{4+}$ state.\cite{xieh}

\subsection{Current-Voltage Characteristics}
Figures 2(a), 2(b), and 2(c) show the room temperature $I$-$V$ data of LSMO/SMO/LSMO trilayer, deposited on STO, LAO, and Si substrates, respectively. A voltage sweeping of range (0 $\rightarrow$ -20 $\rightarrow$ 0 $\rightarrow$ +20 $\rightarrow$ 0) V has been used for the initial $I$-$V$ measurements. The $I(V)$ for the films deposited on STO and LAO substrates shows a symmetric and continuous increase. However, the current in LAO film is one order lower than the film with STO substrate. A nonlinearity in data is evident which becomes more prominent in LAO substrate. A close inspection reveals that the nonlinearity in $I(V)$ occurs at low voltage regime ($<$ 10 V) while in higher voltage regime the $I(V)$ shows almost linear behavior. This nonlinearity is likely due to space charge limited conduction (SCLC) near to the electrode(s) or the creation of Schottky barriers. The enhanced nonlinearity in LAO films is likely due to an interface-induced effect where a comparatively large strain in with LAO substrate has an impact on charge transport.

Figure 2(c) displays the $I$–$V$ characteristics of the LSMO/SMO/LSMO heterostructure on a Si substrate which shows stark contrast from earlier two films with oxide substrates i.e., STO and LAO. As evident, the current has further decreased, and the $I(V)$ is largely asymmetric both in terms of magnitude and shape. However, distinct RS with hysteric $I(V)$ is observed on both polarity of applied voltages. It is further noticed that switching from HRS to LRS (set) and LRS to HRS (reset) occurs at different voltages on negative and positive cycle, where the transition happens at much lower voltage on the negative cycle. Nonetheless, this asymmetric $I(V)$ indicates the influence of Schottky type interfacial barrier potential on the charge transport. 

\begin{figure}
	\centering
		\includegraphics[width=8.5cm]{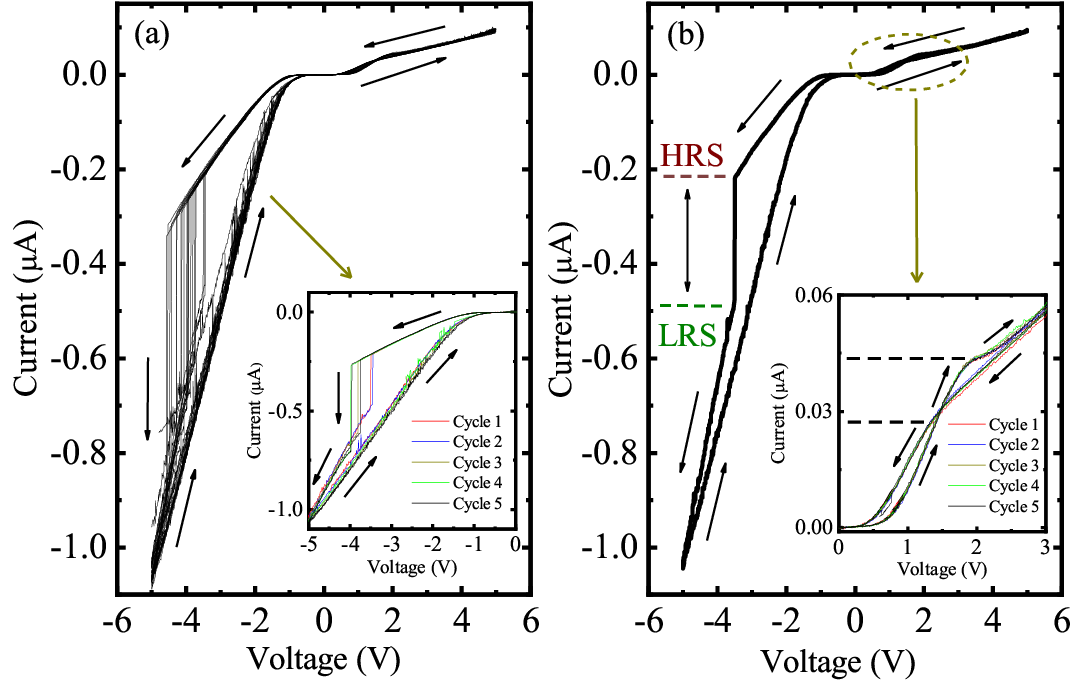}
	\caption{(a) shows the $I$-$V$ characteristics of LSMO/SMO/LSMO film on Si(100) substrate for successive 21 cycles with applied voltage range $\pm$5 V. The direction of voltage cycle is indicated the arrows. Inset shows a magnified view of $I$-$V$ data for initial five cycles in negative voltage range demonstrating a clear RS behavior. (b) shows the representative second cycle showing HRS and LRS state and RS in low negative voltage side. Inset shows a magnified view of positive voltages for initial five cycles showing a consistent anomaly in voltage range between (0.5 - 2.5) V where a crossed double-loop is observed.}
	\label{fig:Fig4}
\end{figure}

For a detailed understanding, the $I(V)$ data have been collected with different voltage sweeping ranges i.e., $\pm$20 V, $\pm$10 V, and $\pm$5 V, as presented in Figs. 3(a), 3(b), and 3(c), respectively. With decreasing voltage range, the RS behavior at high voltage in positive cycle is missed out, however, a close inspection reveals an anomaly at low voltage regime in positive cycle (discussed in Fig. 4(b)). As evident in Fig. 3(c), the RS from LRS to HRS is also absent in negative decreasing voltage for $\pm$5 V. Further, the representative plot in Fig. 3(c) ($\pm$5 V) shows $I(V)$ is very quite asymmetric where the current in negative cycle is around one order higher, compared to positive cycle. Nonetheless, the overall current magnitude as well as the amount of jump is quite small, where this low current is considered to be suitable for device scalability.

\begin{figure}
	\centering
		\includegraphics[width=7cm]{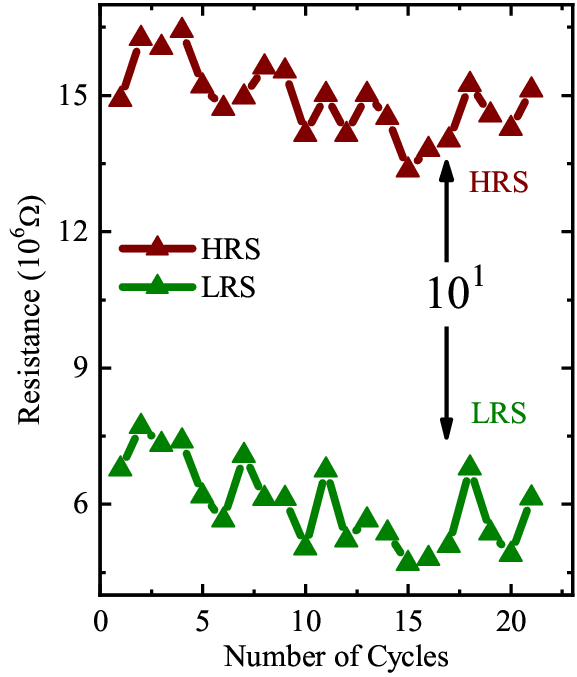}
	\caption{Resistance values at HRS and LRS are shown with voltage sweeping cycle number.}
	\label{fig:Fig5}
\end{figure}

The typical $I$–$V$ characteristics of present film with several voltage cycles ($\pm$5 V) are shown in Fig. 4(a), where a magnified view of $I(V)$ data with negative voltages is shown in inset for successive initial five cycles. On increasing voltage in negative cycle, there is a sudden increase in current for every cycle within voltage range ($|$3.5$|$ - $|$4.0$|$) V, while the current shows a continuous decrease on withdrawal of negative voltage (in contrast with sweeping range $\pm$20 and $\pm$10 V shown in Fig. 3). As evident in inset, the jump in current is mostly consistent. For a better understanding of the $I$–$V$ behavior, we have plotted a representative $I(V)$ in Fig. 4(b) with full voltage sweep range where the direction of voltage sweep is indicated by the arrows in figure. In negative voltage cycle, a RS from HRS to LRS is seen with change in current, $\Delta I$ $\sim$ 0.15 $\mu A$. Inset of Fig. 4(b) shows the magnified view of $I(V)$ data of positive voltages for successive five cycles. Interestingly, a close inspection in $I(V)$ shows an anomaly with crossed double-loops in positive voltage cycle, where $I(V)$ increases/decreases with higher slope. 

\begin{figure}
	\centering
		\includegraphics[width=8.5cm]{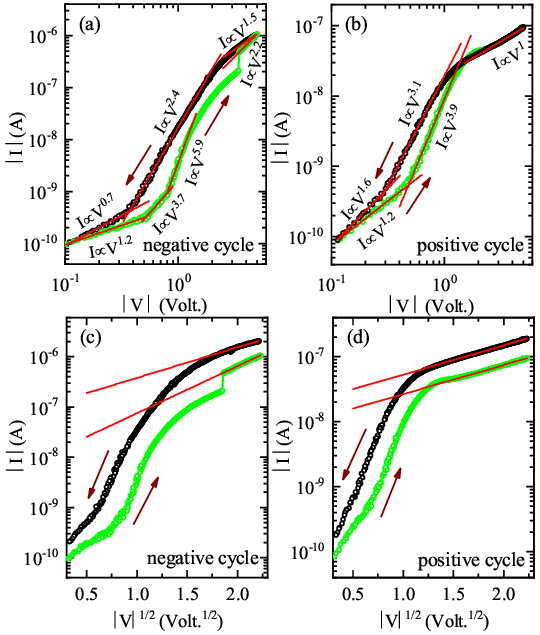}
	\caption{(a) and (b) show the room temperature $I$-$V$ data in $\log$-$\log$ scale for Si(100) based LSMO/SMO/LSMO multilayer with negative and positive applied voltages, respectively where the voltage is scanned in range $\pm$5 V. The red lines are due to fit with $I$ $\propto$ $V^{n}$ for both positive and negative applied voltages. (c) and (d) show the $I$-$V$ data in form of $\log |I|$ vs $|V|^{1/2}$ for both positive and negative voltage regimes, where the data for return cycle has been shifted vertically for clarity.}
	\label{fig:Fig6}
\end{figure}

The observed RS behavior is rather robust against the voltage sweeping. Fig. 5 shows the resistance values at HRS and LRS for successive 21 cycles. It is evident in figure that the resistance values in both states are quite consistent over the voltage cycles, and the ratio between the resistances of HRS and LRS is around 10. Such resistance ratios are commonly observed in oxide-based RS devices, as also reported for Ta$_2$O$_{5-x}$/TaO$_{2-x}$ and La$_{0.7}$Sr$_{0.3}$MO$_3$/SrRuO$_3$ based systems.\cite{mjlee3,snjama} This consistency is in favor of good quality film and interface.

\begin{figure*}
	\centering
		\includegraphics[width=15cm]{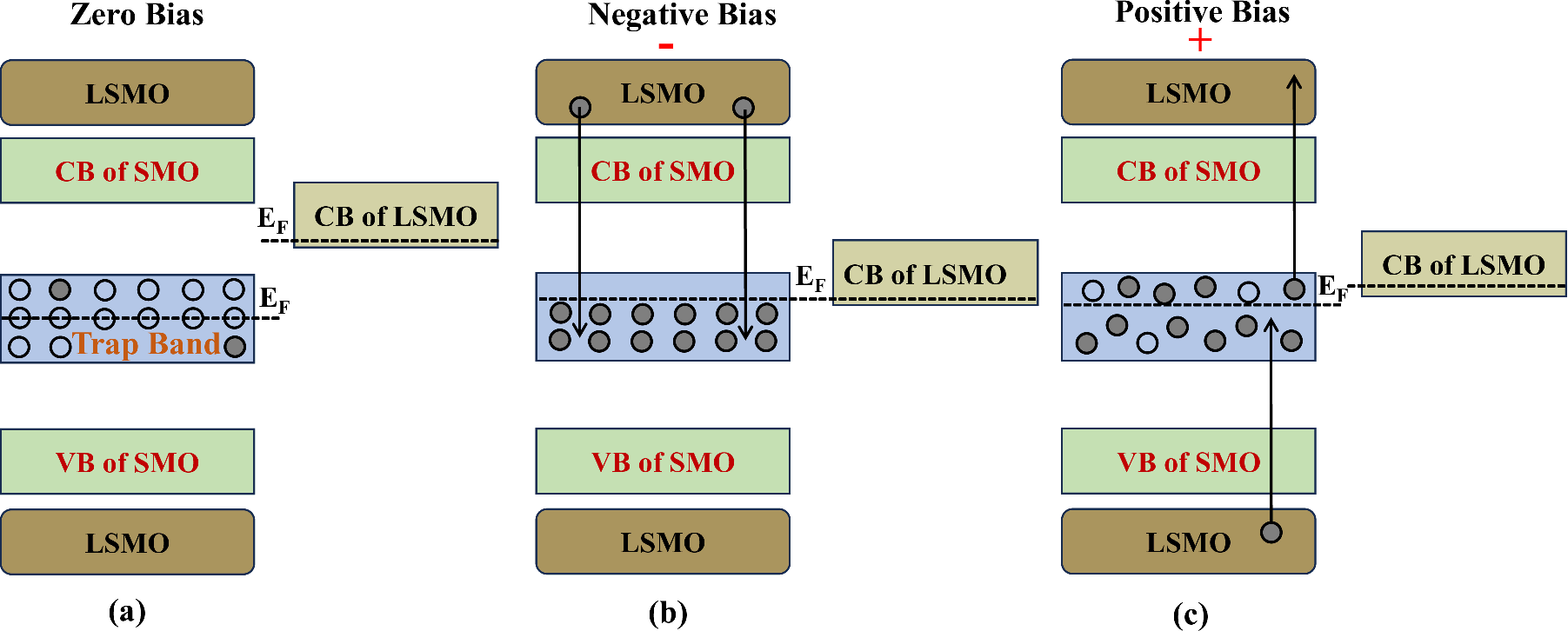}
	\caption{(a) shows the schematic band diagram of LSMO/SMO/LSMO in zero bias voltage. Trap band represents the defect levels in SMO which is energetically distributed around its Fermi level ($E_F$) while CB and VB represent the conduction band and valence band, respectively. A minimal charge flow is expected due to mismatch of $E_F$ between SMO and LSMO. (b) shows the filling of Trap band in SMO with negative bias voltage that results in a matching of $E_F$ and flowing of charges. Depletion and simultaneous filling of Trap band with positive bias voltage has been shown in (c).}
	\label{fig:Fig7}
\end{figure*}

To understand the charge conduction mechanism in present LSMO/SMO/LSMO multilayer deposited on Si substrate, the $I(V)$ data are presented in $\log$-$\log$ scale. Figs. 6(a) and 6(b) show the $|I|$ vs $|V|$ plots for negative and positive applied voltages, respectively. As evident in figures, the $I$ $\propto$ $V^n$ relation is fitted in low voltage regime with the exponent $n$ mentioned in respective figures (red solid lines). For negative cycle, with increasing voltage till $|V|$ $\sim$ 0.5 V, a $n$ = 1.2 suggests an Ohmic-like conduction. However, a large value of $n$ i.e., 3.7 and 5.9 is observed in voltage range (0.5 - 0.8) and (0.8 - 1.2) V, respectively. With further increase in voltage, a signature of RS is observed at $|V|$ $\sim$ 3.4 V while $n$ = 2.2 is observed in high voltage (3.4 - 5.0) regime. The decreasing voltage in negative cycle shows similar Ohmic and non-Ohmic type $I(V)$ behavior ($n$ = 1.5, 2.4 and 0.7) in different voltage regimes, but without any RS behavior (Fig. 6(a)). For positive applied voltages (Fig. 6(b)), we observe similar $I$ $\propto$ $V^n$ is followed in different voltage regimes where both Ohmic and non-Ohmic type conduction ($n$ nearly 1 and above) are evident. Given that the observed $I(V)$ for present heterostructure is highly asymmetric with respect to voltage polarity, the charge conduction is further examined to check the Schottky-type emission. For this, the $I(V)$ data are plotted in form of $\log |I|$ vs $|V|^{1/2}$ for both negative and positive voltage cycles in Figs. 6(c) and 6(d), respectively. As evident in figures, a linear behavior is observed in high field regime suggesting a Schottky-type conduction.

As evident in Fig. 6, the present LSMO/SMO/LSMO multilayer shows voltage-driven different conduction mechanism where LSMO and SMO are characterized with different electronic concentration and mobility. The stiochiometric LSMO has Mn$^{3+}$ (3$d^4$) and Mn$^{4+}$ (3$d^3$) ionic content around 67 and 33\%, respectively, while SMO has only Mn$^{4+}$ ions. The electronic configurations of Mn$^{3+}$ and Mn$^{4+}$ are $t_{2g}^3$,$e_g^1$ and $t_{2g}^3$,$e_g^0$, where $t_{2g}$ and $e_g$ electrons are considered to be localized and itinerant electrons, respectively. Therefore, LSMO has comparatively higher electron density and exhibits metallic character. The current-voltage behavior ($I$ $\propto$ $V^n$), particularly in configuration of insulating medium sandwiched between conducting electrodes (M-I-M), has been described by various models for different values of $n$ where the focus has been given to both intermediate insulating medium as well as conductor-insulator interface(s).

For an Ohomic contact in trap-free insulator, the $I(V)$ characteristics is mainly described by Ohm's law ($n$ = 1) in low voltage regime where the injected charge carrier concentration is less than the intrinsic carrier concentration of the studied material. At higher voltages with larger injection of charge carriers, the space charges near to metal-insulator interface play a crucial role in deciding the conduction mechanism. In case of trap-free space charge limited conduction (SCLC), the current density ($J$) is primarily described by the Mott-Gurney (MG) law: $J$ $\propto$ $V^2/d^3$ where $d$ is the electrode separation.\cite{rosesclc,masclc,ybzhus} The trap-free SCLC ($n$ = 2), has often been discussed in many M-I-M systems,\cite{ndas,kmkim,xia,yliu,acsomega,hchoi} is mainly related to the behavior of the sandwiched component and occurs in materials with relatively low charge mobility when the injected charge carrier concentration surpasses its intrinsic concentration. However, the charge conduction is largely modified in presence of trap/defect sites that capture and release the charge carriers. These trap states have exponential distribution of energy levels, $N(E)$ = $N_t/\left(k_BT_t\right)\exp\left[\left(E-E_c\right)/k_BT_t\right]$, where $k_B$ is the Boltzmann's constant, $N_t$ is the trap density, $T_t$ is the characteristic trap temperature, $E_c$ is the band edge energy.\cite{ybzhus,ysangs} These traps usually stay within the band-gap of insulator where the charge conduction rather happens through tunneling without involving the conduction band. In this trap-limited SCLC regime, the charge conduction follows the Mark-Helfrich (MH) law \cite{ybzhus,ysangs,markpsc,cchuasc} with $J$ $\propto$ $V^{l+1}$/$d^{2l+1}$, where $l$ (= $T_t$/$T$) signifies the ratio between distribution of traps to the free carriers. In trap-limited SCLC region, $l$ is usually $>$ 1 which suggests $n$ to be higher than 2.\cite{yliu,acsomega,hchoi} In fact, a generalized form of $J$ $\propto$ $V^{\alpha}/d^{\beta}$ has been discussed for different kinds of materials.\cite{ysangs} With the increasing electric field, the trap sites will show progressive filling where the sites will be completely filled on reaching the critical voltage. Given that oxide materials, particularly in form of films, usually have oxygen vacancies which act as trap sites, hence the trap-limited SCLC is quite expected in these materials.

In present heterostructure (LSMO/SMO/LSMO), the intermediate SMO is an antiferromagnetic insulator with a band-gap around 1 eV, while LSMO electrodes are ferromagnetic metal.\cite{urushibara,sondena} The schematic of band diagram of LSMO/SMO/LSMO system as well as band filling with bias voltage are shown in Fig. 7. It is fair to assume that the Fermi level in SMO lies near the middle of gap coexisting with the trap levels, while the Fermi level in LSMO stays well within the conduction band. The exact location of Fermi level, however, depends on various factors such as temperature, doping level, stoichiometry, etc. In zero bias voltage, the mismatch of Fermi levels will not allow smooth charge conduction between SMO and LSMO, resulting in HRS (Fig. 7(a)). In fact, in low voltage regime we observe a nearly Ohmic conduction ($n$ $\sim$ 1 in Figs. 6(a) and 6(b)). On application on sufficient negative voltage, the charge carriers will be injected from LSMO to SMO layer, thus filling the trap levels in SMO (Fig. 7(b)). This is represented by a trap-limited SCLC with $n$ $>$ 2. Once the bias voltage is raised to the critical value, or the carrier density in trap states reaches the critical limit (Mott limit), a quasi continuum state is formed in trap levels. This gives a matching of Fermi level of SMO and LSMO, hence a RS is realized from HRS to LRS. Above the RS switching voltage (3.4 - 5.0) V, we get $n$ nearly 2, suggesting a trap-free SCLC behavior. A trap-free SCLC, above the trap-limited SCLC, in high voltages has been shown for ITO/TaO$_x$/NiO$_x$/Al, where the use of additional NiO$_x$ thin layer has shown a better control over the RS behavior.\cite{ybzhus} On withdrawal of negative voltage, reverse phenomena i.e., trap-free, trap-limited and Ohomic conduction are observed. Here, it can be noted that for voltage sweeping up to -5V, RS is not observed with the withdrawal of voltage. However, the RS from LRS to HRS has been observed on return voltage for voltage sweeping up to -20 and -10 V (Figs. 3(a) and 3(b)).

With the positive bias voltage at top electrode (Fig. 7(c)), charge carriers will deplete the trap states. In present multilayer, however, the metallic LSMO being the bottom electrode the charges will be pulled from bottom LSMO to SMO. This results in similar RS behavior with positive bias, though the RS occurs at higher voltage because of reduced electric field. In fact, we have observed the both RS, transition from HRS to LRS and reverse, in same voltage polarity (see Fig. 3(a)). As a charge conduction type, we have successively observed nearly Ohmic ($n$ = 1.2 - 1.6), trap-limited SCLC ($n$ = 3.1 - 3.9) and Ohmic conduction ($n$ = 1) for both increasing and decreasing positive voltages (Fig. 6(b)). Here, note that we observe trap-free SCLC and Ohmic charge transport in high voltages (above trap-limited SCLC) for negative and positive bias voltages, respectively. Further, the exponent $n$ comes out comparatively low during positive voltage scan in both intermediate and high voltage regime. These indicate a reduced charge injection during positive bias because bottom LSMO electrode experiences a diminished electric field. The signature of Schottky-type emission in Figs. 6(c) and 6(d) implies that the M-I interface barrier also contribute to the charge conduction in high voltages. Here, we note that the underlying substrate plays a dominant role for RS in present systems. An absence of RS in films with oxide substrates (STO and LAO) may be due to a charge transfer at LSMO/substrate interface or a different strain and lattice orientation at interface (Fig. 1). Nonetheless, here we categorically show a substrate controlled RS behavior in oxide heterostructures which would be quite useful for memory and memristive devices.  
	
\section{Conclusion}
In conclusion, epitaxial La$_{0.67}$Sr$_{0.33}$MnO$_3$/SrMnO$_3$/ La$_{0.67}$Sr$_{0.33}$MnO$_3$ trilayer films have been successfully deposited using the pulsed laser deposition (PLD) method, with a total thickness of about 120 nm on different substrates, including SrTiO$_3$(100), LaAlO$_3$(100), Si(100). The films have a good crystalline structure, where its epitaxial growth has been confirmed by x-ray diffraction (XRD) measurements. While the atomic force microscopy (AFM) has shown smooth film surface with minimal roughness, the Raman spectroscopy shows usual peaks due to mixed Mn$^{3+}$/Mn$^{4+}$ ionic state. The room temperature $I$-$V$ data with multilayers grown on oxide SrTiO$_3$ and LaAlO$_3$ substrates exhibit regular with slight nonlinearity at low applied voltages. In contrast, a notable asymmetry and significant resistance switching (RS) behavior are observed in $I$-$V$ data with multilayer deposited on Si substrate. Our analysis reveals that the charge conduction in low voltage regime follows Ohmic and SCLC type mechanism while in high voltage regime Schottky behavior is observed. This substrate driven tunability of $I$-$V$, where the lattice strain and material chemistry at interface plays crucial role, is rather interesting and shows its potential for future memory and memristive devices.

\section{Acknowledgment}
We thank AIRF, JNU for Atomic Force Microscope (AFM) measurements, UPE-II for the deposition chamber, and SERB-DST for supporting the Excimer Pulse Laser. We also acknowledge Dr. Ajay Shukla, National Physical Laboratory, Delhi for the Raman measurements and Dr. Uday Deshpande, UGC-DAE Indore for XPS measurements and Mr. Sachin Dabral for recording XPS data. A. G. A. Rahman expresses gratitude to CSIR, India for the financial assistance.


\begin{thebibliography}{}
\bibitem{hiatt} Hiatt, W. R.; Hickmott, T. W. Bistable Switching in Niobium Oxide diodes. D Appl. Phys. Lett. \textbf{1965}, 6, 106.
\bibitem{waser} Waser, R.; Dittmann, R.; Staikov, G.; Szot, K. Redox-Based Resistive Switching Memories-Nanoionic Mechanisms, Prospects, and Challenges. Adv. Mater. \textbf{2009}, 21, 2632.
\bibitem{shang} Da-Shan, S.; Ji-Rong, S.; Bao-Gen, S.; Matthiasb, M. Resistance switching in oxides with inhomogeneous conductivity. Chin. Phys. B \textbf{2013}, 22, 067202.
\bibitem{pan} Pan, F.; Gao, S.; Chen, C.; Song, C.; Zeng, F. Recent progress in resistive random access memories: Materials, switching mechanisms, and performance. Mater. Sci. Eng. R \textbf{2014}, 83, 1-59.
\bibitem{baikalov} Baikalov, A.; Wang, Y.Q.; Shen, B.; Lorenz, B.; Tsui, S.; Sun, Y.Y.; Xue, Y.Y.; Chu, C.W. Field-driven hysteretic and reversible resistive switch at the Ag–Pr$_{0.7}$Ca$_{0.3}$MnO$_3$ interface. Appl. Phys. Lett. \textbf{2003}, 83, 957–959.
\bibitem{rwaser} Waser, R.; Aono, M. Nanoionics-based resistive switching memories. Nat. Mater. \textbf{2007}, 6, 833–840.
\bibitem{nian} Nian, Y.B.; Strozier, J.; Wu, N.J.; Chen, X.; Ignatiev, A. Evidence for an Oxygen Diffusion Model for the Electric Pulse Induced Resistance Change Effect in Transition-Metal Oxides. Phys. Rev. Lett. \textbf{2007}, 98, 146403.
\bibitem{liu} Liu, S.Q.; Wu, N.J.; Ignatiev, A. Electric-pulse-induced reversible resistance change effect in magnetoresistive films Available. Appl. Phys. Lett. \textbf{2000}, 76, 2749– 2751.
\bibitem{chudnovskii} Chudnovskii, F.A.; Odynets, L.L; Pergament, A.L.; Stefanovich, G.B. Electroforming and Switching in Oxides of Transition Metals: The Role of Metal–Insulator Transition in the Switching Mechanism. J. Solid State Chem. \textbf{1996}, 122, 95–99.
\bibitem{rossel} Rossel, C.; Meijer, G.I.; Bremaud, D.; Widmer, D. Electrical current distribution across a metal–insulator–metal structure during bistable switching. J. Appl. Phys. \textbf{2001}, 90, 2892–2898.
\bibitem{kwon} Kwon, D.H.; Kim, K.M.; Jang, J.H.; Jeon, J.M.; Lee, M.H.; Kim, G.H.; Li, X.S.; Park, G.S.; Lee, B.; Han, S.; Kim, M.; Hwang, C.S. Atomic structure of conducting nanofilaments in TiO2 resistive switching memory. Nat. Nanotechnol. \textbf{2010}, 5, 148–153.
\bibitem{lee} Lee, M.J.; Han, S.; Jeon, S.H.; Park, B.H.; Kang, B.S.; Ahn, S.E.; Kim, K.H.; Lee, C.B.; Kim, C.J.; Yoo, I.K.; Seo, D.H.; Li, X.S.; Park, J.B.; Lee, J.H.; Park, Y. Electrical Manipulation of Nanoﬁlamentsin Transition-Metal Oxides for Resistance-Based Memory. Nano Lett. \textbf{2009}, 9, 1476–1481.
\bibitem{tsui} Tsui, S.; Baikalov, A.; Cmaidalka, J.; Sun, Y.Y.; Wang, Y.Q.; Yue, Y.Y.; Chu, C.W.; Chen, L.; Jacobson, A. Field-induced resistive switching in metal-oxide interfaces. J. Appl. Phys. Lett. \textbf{2004}, 85, 317–319.
\bibitem{odagawa} Odagawa, A.; Kanno, T.; Adachi, H. Transient response during resistance switching in Ag/Pr$_{0.7}$Ca$_{0.3}$MnO$_3$/Pt thin films. J. Appl. Phys. \textbf{2006}, 99, 016101.
\bibitem{sawa} Sawa, A. Resistive switching in transition metal oxides. Mater. Today \textbf{2008}, 11, 28-36.
\bibitem{gao} Gao, T.T.; Tan, T.T.; Liu, Z.T. Effects of Film Thickness and Ar/O$_2$ Ratio on Resistive Switching Characteristics of HfO$_x$-Based Resistive-Switching Random Access Memories. Chin. Phys. Lett. \textbf{2015}, 32, 016801.
\bibitem{szot} Szot, K.; Speier, W.; Gihlmayer, G.; Waser, R. Switching the electrical resistance of individual dislocations in single-crystalline SrTiO$_3$. Nat. Mater. \textbf{2006}, 5, 312-320.
\bibitem{lai} Lai, X.B.; Wang, Y.H.; Shi, X.L.; Li, D.Y.; Liu, B.Y.; Wang, R.M.; Zhang, L.W. Bipolar Resistive Switching in Epitaxial Mn$_3$O$_4$ Thin Films on Nb-Doped SrTiO$_3$ Substrates. Chin. Phys. Lett. \textbf{2016}, 33, 067202.
\bibitem{guo} Guo, Z.; Li, M.Q.; Huang, X.J. Cation Exchange Synthesis and Unusual Resistive Switching Behaviors of Ag$_2$Se Nanobelts. Small \textbf{2015}, 11, 6285.
\bibitem{zhou} Zhou, J.W.; Li, T.R.; Zhang, D.; Ren, B.; Zhang, S.W.; Huang, J.; Zhang, J.M.; Wang, L.; Jiang, Y.C.; Gao, J.; Wang, L.J. Abnormal bipolar resistive switching behavior in carbon-iron composite films with different thicknesses. Vacuum \textbf{2017}, 135, 115-120. 
\bibitem{chen} Chen, C.; Yang, Y.C.; Zeng, F.; Pan, F. Bipolar resistive switching in Cu/AlN/Pt nonvolatile memory device. Appl. Phys. Lett. \textbf{2010}, 97, 083502.
\bibitem{kim} Kim, T.W.; Zeigler, D.F.; Acton, O.; Yip, H.L.; Ma, H.; Jen, A.K.Y. All-Organic Photopatterned One Diode-One ResistorCell Array for Advanced Organic Nonvolatile MemoryApplications. Adv. Mater. \textbf{2012}, 24, 828-833.
\bibitem{khare} Khare, A.; Shin, D.; Yoo, T.S.; Kim, M.; Kang, T.D.; Lee J.; Roh S.; Jung I.; Hwang J.; Kim S.W.; Noh T.W.; Ohta H.; Choi W.S.; Topotactic Metal–Insulator Transition in Epitaxial SrFeO$_x$ Thin Films. Adv. Mater. \textbf{2017}, 29, 1606566.
\bibitem{kubicek} Kubicek, M.; Bork, A.H.; Rupp, J.L.M. Perovskite oxides – a review on a versatile material class for solar-to-fuel conversion processes. J. Mater. Chem. A \textbf{2017}, 5, 11983.
\bibitem{hu} Hu, W.J.; Hu, L.; Wei, R.H.; Tang, X.W.; Song, W.H.; Dai, J.M.; Zhu, X.B.; Sun, Y.P. Nonvolatile Resistive Switching and Physical Mechanism in LaCrO$_3$ Thin Films. Chin. Phys. Lett. \textbf{2018}, 35, 4-047301.
\bibitem{urushibara} Urushibara, A.; Moritomo, Y.; Arima, T.; Asamitsu, A.; Kido, G.; Tokura, Y. Insulator-metal transition and giant magnetoresistance in La$_{1-x}$Sr$_x$MnO$_3$. Phys. Rev. B. \textbf{1995}, 51, 14103.
\bibitem{hemberger} Hemberger, J.; Krimmel, A.; Kurz, T.;  Nidda, H.A.K.v.; Ivanov, V. Yu.; Mukhin, A.A.; Balbashov, A.M.; Loidl, A. Structural, magnetic, and electrical properties of single-crystalline La$_{1-x}$ Sr$_x$MnO$_3$ (0.4 $<x<$0.85). Phys. Rev. B \textbf{2002}, 66, 094410.
\bibitem{mjlee} Lee, M.J.; Lee, C.B.; Lee, D.; Lee, S.R.; Chang, M.; Hur, J.H.; Kim, Y.B.; Kim, C.J.; Seo, D.H.; Seo, S.; Chung, U.I.; Yoo, I.K.; Kim, K. A fast, high-endurance and scalable non-volatile memory device made from asymmetric Ta$_2$O$_{5-x}$/TaO$_{2-x}$ bilayer structures. Nat. Mater. \textbf{2011}, 10, 625–630.
\bibitem{ulrichs} Ulrichs, H.; Demidov, V.E.; Demokritov, S.O.; Urazhdin, S. Parametric excitation of eigenmodes in microscopic magnetic dots. Phys. Rev. B \textbf{2011}, 84, 094401.
\bibitem{kharkwal} Kharkwal, K.C.; Chaurasia, R.; Pramanik, A.K. Unusual exchange bias in Sr$_2$FeIrO$_6$/La$_{0.67}$Sr$_{0.33}$MnO$_3$ multilayer. J. Phys. Condens. Matter \textbf{2019}, 31, 13LT02. 
\bibitem{granado} Granado, E.; Moreno, N.O.; Garcia, A.; Sanjurjo, J.A.; Rettori, C.; Torriani, I.; Oseroff, S.B.; Neumeier, J.J.; McClellan, K.J.; Cheong, S.W.; Tokura, Y. Phonon Raman scattering in R$_{1-x}$A$_x$MnO$_{3+\delta}$ (R=La,Pr; A =Ca,Sr). Phys. Rev. B. \textbf{1998}, 58, 11435.
\bibitem{sacchetti} Sacchetti, A.; Baldini, M.; Postorino, P.;  Martin, C.; Maignan, A.; Raman spectroscopy on cubic and hexagonal SrMnO$_3$. J. Raman Spectrosc. \textbf{2006}, 37, 591–596. 
\bibitem{behera} Behera, B.C.; Ravindra, A.V.; Padhan, P.; Prellier, W.; Raman spectra and magnetization of all-ferromagnetic superlattices grown on (110) oriented SrTiO$_3$. Appl. Phys. Lett. \textbf{2014}, 104, 092406.
\bibitem{rlijin} Li, R.; Jin, C.; Bai, H.; Sign reversal and symmetry change of anisotropic magnetoresistance in antiferromagnetic LSMO films. J. Appl. Phys. \textbf{2024}, 136, 043901. 
\bibitem{anne} Annesea, E.; Morib, T.J.A.; P. Schiob, P.; Bonfimc, R.P.F.; Sallesd, B.R.; Sernad, J.D.P.; Cezarb, J.C. Magnetic properties of La$_{0.67}$Sr$_{0.33}$MnO$_3$ epitaxial thin films grown on SrTiO$_3$(001). J. Magn. Magn. Mater. \textbf{2020}, 507, 166812.
\bibitem{ren} Ren, R.; Ren, Y.; Li, X.; Wen, L.; Chen, S.; Wu, Z.; Xu, R. Transport dynamics analysis in ferromagnetic heterojunction using Raman spectroscopy and magnetic force microscopy. Prog. Nat. Sci: Mat. Int. \textbf{2016}, 26, 173–176.
\bibitem{aga} Rahman, A.G.A.; Patel, R.K.; Sachin, S.; Kumar, H.; Nath, C.; Chakravarty, S.; Manna, S.; Pramanik, A.K. Tuning of magnetic and transport behavior in La$_{0.67}$Sr$_{0.33}$MnO$_3$/Pr$_2$Ir$_2$O$_7$ bilayer: Possible role of interfacial Dzyaloshinskii–Moriya interactions. Surfaces and Interfaces \textbf{2025}, 56, 105447.
\bibitem{yangconf} Yang, H. B. The Structural and Morphology of La$_{0.6}$Sr$_{0.4}$MnO$_3$ Thin Films Prepared by Pulsed Laser Deposition. MATEC Web of Conferences.\textbf{2016}, 44, 02035.
\bibitem{xpsbook} Wagner, C. D. et al. Handbook of X-ray Photoelectron Spectroscopy [Wagner,C. D., Riggs, W. M., Davis, L. E., Moulder, J. F. (ed.)] (Eden Prairie, Minnesots \textbf{1979}, 55344).
\bibitem{burroughs} Burroughs, P.; Hammett, A.; Orchard, A. F.; Thornton, G. Satellite Structure in the X-Ray Photoelectron Spectra of some Binary and Mixed Oxides of Lanthanum and Cerium. J. Chem. Soc. Dalton Trans. \textbf{1976}, 1686. 
\bibitem{viswanathan} Viswanathan, B.; Madhavan, S.; Swamy, C. S. Charge Transfer Satellites in X-Ray Photoelectron Spectra of La$_2$Cu0$_4$. Phys. Status Solidi B \textbf{1986}, 133, 629.
\bibitem{vasquez} Vasquez, R. P. X-ray Photoemission Measurements of La$_{1-x}$Ca$_x$CoO$_3$ (x = 0, 0.5). Phys. Rev. B. \textbf{1996}, 54, 14938.
\bibitem{bertaccor} Bertacco, R.; Contour, J. P.; Barthelemy, A.; Olivier, J. Evidence for strontium segregation in La$_{0.7}$Sr$_{0.3}$MnO$_3$ thin films grown by pulsed laser deposition: consequences for tunnelling junctions. Surface Science \textbf{2002} 511, 366.
\bibitem{chaluvadi} Chaluvadi, S. K.; Polewczyk, V.; Petrov, A. Y.; Vinai, G.; Braglia, L.; Diez, J. M.; Pierron, V.; Perna, P.; Mechin, L.; Torelli, P.; Orgiani, P. Electronic Properties of Fully Strained La$_{1-x}$Sr$_x$MnO$_3$ Thin Films Grown by Molecular Beam Epitaxy (0.15 $\leq$ x $\leq$ 0.45). ACS Omega \textbf{2022}, 7, 14571.
\bibitem{xieh} Xie, H.; Huang, H.; Cao, N.; Zhou, C.; Niu, D.; Gao, Y. Effects of annealing on structure and composition of LSMO thin films. Physica B \textbf{2015}, 477, 14.
\bibitem{mjlee3} Lee, M-J; Lee, C. B.; Lee, D.; Lee, S. R.; Chang, M.; Hur, J. H.; Kim, Y-B.;  Kim, C-J.; Seo, D. H.; Seo, S.; Chung, U-In.;  Yoo, In-Kyeong.; Kim, K. A fast, high-endurance and scalable non-volatile memory device made from asymmetric Ta$_2$O$_{5-x}$/TaO$_{2-x}$ bilayer structures Nat. Mater. \textbf{2011}, 10, 625. 
\bibitem{snjama} Jammalamadaka,  S. N.; Vanacken, J.; Moshchalkov, V. V., Resistive switching in ultra-thin La$_{0.7}$Sr$_{0.3}$MnO$_3$/SrRuO3 superlattices  Appl. Phys. Lett. \textbf{2014}, 105, 033505.
\bibitem{rosesclc} Rose, A. Space-charge-limited currents in solids. Physical Review \textbf{1955}, 97, 1538.
\bibitem{masclc} Lampert, M. A.; Mark, P. Current injection in solids (Academic, New York) \textbf{(1970)}.
\bibitem{ybzhus} Zhu, Y. B.; Zheng, K.; Wu, X.; Ang, L. K. Enhanced stability of filament-type resistive switching by interface engineering. Sci. Rep. \textbf{2017}, 7, 43664.
\bibitem{ysangs} Ang, Y. S.; Zubair, M.; Ang, L. K. Relativistic space-charge-limited current for massive Dirac fermions. Phys. Rev. B \textbf{2017}, 95, 165409. 
\bibitem{markpsc} Mark, P.; Helfrich, W. Space-charge-limited currents in organic crystals. J. Appl. Phys. \textbf{1962} 33, 205.
\bibitem{cchuasc} Chua, C.; Ang, Y. S.;  Ang, L. K. Tunneling injection to trap-limited space-charge conduction for metal-insulator junction. Appl. Phys. Lett. \textbf{2022}, 121, 192109.
\bibitem{ndas} Das, N.; Tsui, S.; Xue, Y.Y.; Wang, Y.Q.; Chu, C.W. Electric-field-induced submicrosecond resistive switching. Phys. Rev. B \textbf{2008}, 78, 235418. 
\bibitem{kmkim} Kim, K.M.; Choi, B.J.; Shin, Y.C.; Choi S.; Hwanga, C.S. Anode-interface localized filamentary mechanism in resistive switching of TiO$_2$ thin films. Appl. Phys. Lett. \textbf{2007}, 91, 012907. 
\bibitem{xia} Xia, Y.; He, W.; Chen, L.; Meng X.; Li, Z. Field-induced resistive switching based on space-charge-limited current. Appl. Phys. Lett. \textbf{2007}, 90, 022907.
\bibitem{yliu} Liu, Y.; Gao, P.; Jiang, X.; Li, L.; Zhang J.; Peng, W. Percolation mechanism through trapping/de-trapping process at defect states for resistive switching devices with structure of Ag/Si$_x$C$_{1-x}$/p-Si Available. J. Appl. Phys. \textbf{2014}, 116, 064505. 
\bibitem{acsomega} Yu, Y.; Ding, Z.; Ren, Y.; Wang, X.; Quan, H.; Jia H.; Jiang, C. Understanding the Resistive Switching Behaviors of Top Electrode (Au, Cu, and Al)-Dependent TiO$_2$-Based Memristive Devices. ACS Omega \textbf{2024}, 9, 24601--24609.
\bibitem{hchoi} Choi, H.J.; Park, S.W.; Han, G.D.; Na, J.; Kim, G.T. Resistive switching characteristics of polycrystalline SrTiO$_3$ films. Appl. Phys. Lett. \textbf{2014}, 104, 242105.
\bibitem{sondena} Sondena, R.; Ravindran, P.; Stolen, S. Electronic structure and magnetic properties of cubic and hexagonal SrMnO$_3$. Phys. Rev. B  \textbf{2006}, 74, 144102.


\end{thebibliography}
\end{document}